\documentclass[aps,prl,preprint,groupedaddress,amsmath,amssymb]{revtex4-2}
\usepackage{etoolbox}

\makeatletter
\patchcmd{\frontmatter@abstract@produce}
  {\vskip200\p@\@plus1fil
   \penalty-200\relax
   \vskip-200\p@\@plus-1fil}
  {}
  {}
  {}
\makeatother

\usepackage{graphicx}

\usepackage{hyperref}
\hypersetup{colorlinks,linkcolor={blue},citecolor={red},urlcolor={blue}}  

\begin{document}

\title{Comment on “Absence of Topological Protection of the Interface States in \texorpdfstring{$\mathbb{Z}_2$}{Z\_2} Photonic Crystals”}

\author{Xing-Xiang Wang}
\affiliation{Research Center for Materials Nanoarchitectonics (MANA), National Institute for Materials Science (NIMS), Tsukuba 305-0044, Japan}
\author{Toshikaze Kariyado}
\affiliation{Research Center for Materials Nanoarchitectonics (MANA), National Institute for Materials Science (NIMS), Tsukuba 305-0044, Japan}
\author{Xiao Hu}
\affiliation{Research Center for Materials Nanoarchitectonics (MANA), National Institute for Materials Science (NIMS), Tsukuba 305-0044, Japan}

\maketitle

In the Letter \cite{xuAbsence2023}, Xu et al. reported that edge modes disappear in the expanded structure of Wu-Hu model 
characterized by $\mathbb{Z}_2$ topological index \cite{wuScheme2015,wuTopological2016}, while appear in the trivial shrunken structure, 
when the edge cuts through the hexagonal unit cell. 
They then concluded that these edge modes are defect modes lacking topological protection. 
Unfortunately, their approach is not justified, rendering the conclusion unsolid. \par

First, discussions in the Letter \cite{xuAbsence2023} on the fate of edge states 
and the attempt to identify the corresponding topology by edge states 
when the edge cuts through the hexagonal unit cell are invalid. 
It is important to note that the unit cell and edge morphology should be tied one by one 
when considering bulk-edge correspondence, and the unit cell used 
for calculating topological index must be kept intact when looking for corresponding edge states. 
Cutting edge in a different way in fact changes the unit cell, which induces a position-dependent phase factor \cite{kariyadoSymmetryprotected2013}
$u_{n,\mathbf{k}}(\mathbf{r}_\alpha)\rightarrow e^{i \zeta_\mathbf{k} (\mathbf{r}_\alpha)} u_{n,\mathbf{k}}(\mathbf{r}_\alpha)$
with $\alpha$ the site index ($U(6)$ gauge in Wu-Hu model with 6 sites), yielding a nonvanishing term in Berry curvature
\begin{equation}
    \Omega_n(\mathbf{k}) 
    \rightarrow \Omega_n(\mathbf{k}) - 
    \nabla_\mathbf{k} \times \sum_{\alpha}
    [\nabla_\mathbf{k} \zeta_\mathbf{k} (\mathbf{r}_\alpha)]u_{n,\mathbf{k}}^*(\mathbf{r}_\alpha) u_{n,\mathbf{k}}(\mathbf{r}_\alpha),
\end{equation}
thus modifies the topology, even though the bulk band structure remains unchanged. \par

\begin{figure}
    \includegraphics[width=16cm]{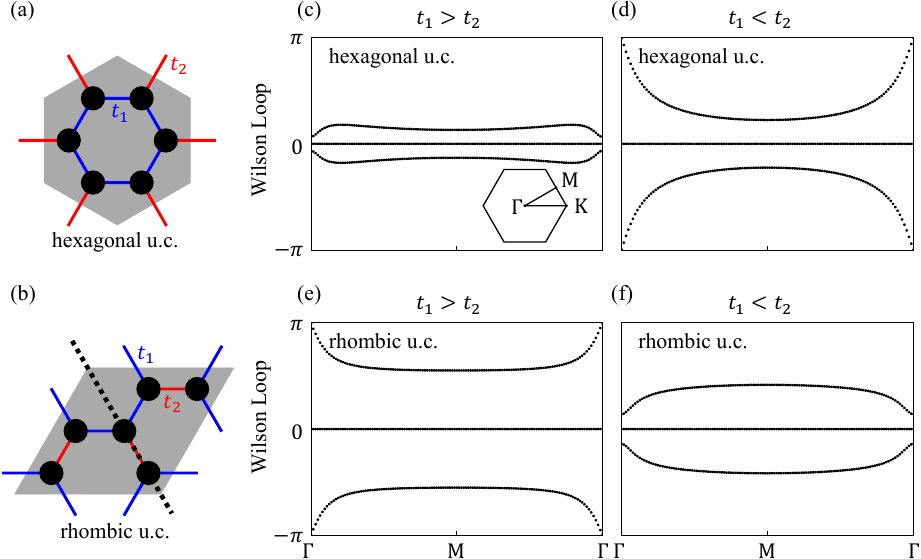}
    \caption{\label{fig:1}(a) and (b) Schematics for the hexagonal unit cell and the rhombic unit cell. 
    Blue and red bonds represent $t_1$ and $t_2$, respectively. 
    Black dashed line in (b) stands for the mirror plane. 
    (c) and (d) Wilson loops for the hexagonal unit cell. 
    (e) and (f) Wilson loops for the rhombic unit cell. 
    The integration direction of the Wilson loops in (e) and (f) is perpendicular to the mirror plane.  
    The first Brillouin zone is given in (c).}
\end{figure}

As a matter of fact, it has been revealed that the system with $t_1>t_2$ (shrunken structure) 
in FIG. 1(a) of the Letter \cite{xuAbsence2023} is characterized by non-zero mirror winding numbers 
when a rhombic unit cell is taken (see Figure. \ref{fig:1}); gapless helical edge states exist 
along the partial-bearded edge compatible with the rhombic unit cell \cite{kariyadoTopological2017,freeneyEdgeDependent2020}. 
We point out that the edge shape in FIG. 2(b) in the Letter \cite{xuAbsence2023} corresponds exactly to the rhombic unit cell, 
which remains intact along the edge whereas the hexagonal unit cell is broken. 
This can also be demonstrated by the Wilson loop spectra as shown in Figure 1, 
which exhibits clearly the dual relation between the two cases of opposite relative strength of $t_1$ and $t_2$, 
and explains fully the results in FIG. 2 in the Letter \cite{xuAbsence2023}. 
The gap opening in the dispersions of edge states in FIG. 2(d) and 2(e) is merely due to the choice of edge direction, 
where the mirror operation mixes the two sublattices \cite{kariyadoTopological2017}.\par

In addition, we note that the topology and corresponding helical edge states 
in the expanded structure of Wu-Hu model have been described successfully 
by  $\pm 2\pi$-winding Wilson loops based on the two-fold pseudospin subspace supported by the $C_2T$ symmetry \cite{palmerBerry2021}. 
Namely, taking pseudospin into account, states in the expanded structure are Wannier obstructed, 
thus topologically nontrivial in accordance with the criterion of Letter \cite{xuAbsence2023}. 
Without closing the band gap and breaking the $C_{6v}$ symmetry, 
this structure can be connected adiabatically to another model of fragile topology \cite{depazEngineering2019}, 
which indicates that the two models share the same topology and accounts for the same helical edge states of the two models \cite{palmerBerry2021}. 
In this sense, consequences of categorization of topology straightforwardly 
based on elementary band representation are not necessarily clear for the associated edge states.\par

Lastly, we cast doubt on the discussion of the Letter \cite{xuAbsence2023} 
that gapping the edge states by including long-range hopping integrals reflects the absence of topological protection. 
Theoretically the topological protection for armchair edges/interfaces is not strict, 
since the sublattice symmetry is missing under the mirror operation mentioned above which opens a gap in the edge/interface states \cite{kariyadoTopological2017}. 
Moreover, the hopping integrals taken for FIG. 3(b) are totally unphysical, 
where sites separated farther have stronger couplings (see Supplementary Information of the Letter \cite{xuAbsence2023}). 
Including more unphysical “perturbations” can change the effective spatial dimension, which eventually damages the topology.  

\begin{acknowledgments}
    This work is partially supported by CREST, JST (Core Research for Evolutionary Science and Technology, Japan Science and Technology Agency) 
    under the grant number JPMJCR18T4. 
\end{acknowledgments}

\bibliography{ref}

\newpage
\end{document}